\documentclass[preprint,pre,amsmath,nofootinbib]{revtex4-2}
\usepackage{amsmath,amssymb,amsfonts,xcolor,graphicx,cases,wasysym}

\makeatletter
\let\old@makecaption=\@makecaption
\usepackage{subcaption}
\let\@makecaption=\old@makecaption
\makeatother
\usepackage{blindtext}

\usepackage{amsmath,hyperref}

\usepackage{dutchcal}

\def\be{\begin{equation}}
\def\ee{\end{equation}}
\def\bea{\begin{eqnarray}}
\def\eea{\end{eqnarray}}
\def\bsn{\begin{subnumcases}}
\def\esn{\end{subnumcases}}

\def\eq#1{(\ref{#1})}
\def\bm{\begin{pmatrix}}
\def\em{\end{pmatrix}}
\def\nn{\nonumber}
\def\bi{\begin{itemize}}
\def\ei{\end{itemize}}

\begin{document}

\title{Jamming transition in an active exclusion process}
\author{Kavita Jain} 
\email{jain@jncasr.ac.in} 
\affiliation{Theoretical Sciences Unit, \\Jawaharlal Nehru Centre for Advanced Scientific Research, Bangalore 560064, India} 
\author{Sakuntala Chatterjee} 
\email{sakuntala.chatterjee@bose.res.in} 
\affiliation{Department of Physics of Complex Systems, S.N. Bose National Center for Basic Sciences, Salt Lake, Sector 3, JD Block, Kolkata 700106, India}

\date{\today} 

\begin{abstract}
Multiple studies on active matter have shown that activity can induce or suppress a phase transition, or modify the critical behavior of a passive system. Here we investigate how activity affects the jamming transition which is a paradigmatic  example of nonequilibrium phase transitions in passive systems. We consider a one-dimensional system of active particles with hardcore interactions and a direction of self-propulsion which can be reversed at a given switching rate. For a class of particle hop rates and infinite switching rate, our model reduces to a passive system which is known to exhibit a transition between a high-density fluid phase and a low-density jammed phase in the stationary state. Using Monte Carlo simulations and a mean field theory, we study how the mean mobility of the particle and the hole cluster distribution vary with density and finite switching rate. Our main result is that activity hinders the formation of jam and can even inhibit it; more precisely, we find that the jamming transition occurs at a critical density that decreases with decreasing switching rate, and at sufficiently small switching rate, the system exists only in the fluid phase. 
\end{abstract}
\maketitle
\clearpage
%

\section{Introduction}

{\color{black}  Active matter refers to a system of interacting particles that consume energy from their environment and produce directed motion \cite{Schweitzer:2003,Bechinger:2016,Ramaswamy:2017}. Natural phenomena across a wide spectrum of length scales, ranging from the concerted flight of a flock of birds \cite{Toner:1995,Toner:1998,Tu:1998,Gregoire:2004,Vicsek:2012} to spatio-temporal patterns in migrating bacterial colonies  \cite{Dellarciprete:2018,Jeckel:2019,Dhar:2022}, can be described by the physics of active matter \cite{Ramaswamy:2010,Marchetti:2013,Cavagna:2014,Vrugt:2026}. The activity, {\it i.e.}, presence of directionality at the single particle level ensures that the detailed balance is broken even at the microscopic level, which makes active matter systems intrinsically out of equilibrium. The activity also gives rise to novel features that are unique to active matter systems and do not have a counterpart in other nonequilibrium systems where activity is absent (passive systems); examples include flocking transition \cite{Vicsek:1995,Solon:2013}, motility-induced phase separation \cite{Tailleur:2008,Cates:2015}, self-assembly \cite{Stenhammar:2015, Adachi:2022,Hahn:2025}, etc.. Conversely, many interesting phenomena, notably phase transitions, that are observed in passive systems may either disappear or get significantly altered upon the introduction of the activity \cite{Tailleur:2008,Thompson:2011,Fily:2012,Solon:2013,Soto:2014,Solon:2015,Siebert:2018,Linden:2019,Frydel:2021,Zhao:2023,Bhowmick:2025,Zhou:2026}. In the present work, we are interested in the latter aspect of the active systems; specifically, we consider a passive system that exhibits a nonequilibrium phase transition in the stationary state and investigate how its phase diagram is affected due to activity.}

In passive systems, a paradigmatic example of nonequilibrium phase transitions is the condensation transition between a low density phase where particle clusters have size of order one and a high density phase in which particles aggregate  into single or few clusters of macroscopic size \cite{Majumdar:1998,Evans:2005}. This transition has been shown exactly in a Zero Range Process (ZRP) for which the exact steady state measure can be obtained in any dimension \cite{Spitzer:1970,Evans:2005}. The canonical process has been generalized in several directions 
\cite{Krug:1996,Jain:2003,Chatterjee:2008,Hirschberg:2009,Thompson:2011,Povolotsky:2013,Evans:2014,Grange:2020} 
and also applied to model various nonequilibrium phenomena \cite{Kaupuzs:2005,Jain:2005,Harris:2005,Chleboun:2010,Priyanka:2016b,Chakraborty:2024}. 
The ZRP can  be mapped to an exclusion process (EP)  \cite{Evans:2005} which describes the movement of hard core particles on a lattice, and has been used to understand traffic flow \cite{Kaupuzs:2005} and to develop hydrodynamic theories \cite{Priyanka:2016a,Chakraborty:2020,Slanina:2022}. 
The condensation transition in ZRP appears as the  {\it jamming transition} in EP in which at low particle densities, one or more   hole clusters of macroscopic size form 
while at high densities, the typical length of hole clusters is of order one. 

Here, we are interested in understanding how the jamming transition in a one-dimensional exclusion process is modified
 when the particles are active. For this purpose, each particle is assigned a direction of self-propulsion which is modeled here as a binary variable, $\eta=\pm 1$ that can switch between the two states at a given rate; in related models  \cite{Soto:2014,Slowman:2016,Das:2025}, such particles are referred to as run-and-tumble particles with $+(-)$ particle allowed to move only to the right (left).   
Depending on its orientation, a particle hops to the neighboring site, provided it is empty, at a rate that  is chosen to be the one for which jamming transition occurs in the passive model \cite{Evans:2005,Priyanka:2016a}.

Our main finding is that activity hinders the formation of jam and can even inhibit it (see Fig.~\ref{mobi}). To understand this result, consider the above model when the particle density is sufficiently low, and the system is in the jammed state with the macroscopic hole cluster bounded by a left mover on the left and right mover on the right. If the particle on (say) the left of the hole cluster switches its sign frequently, before the tumbled particle can hop to the right, its sign may switch back thus stabilizing the hole cluster (passive limit). On the other hand, for sufficiently small switching rates (high activity), long periods may elapse between successive tumblings thus enabling the particle to hop; this would then allow the particles behind the mobile particle to also move (when they have the correct orientation) thus destablizing the jam. In addition to unjamming at low particle densities, we also find that within the fluid phase, the mean mobility of a particle has qualitatively different behavior at low and high switching rates (relative to the density), and the average size of the largest hole cluster varies nonmonotonically with the switching rate. 

\color{black}{In the following section, we define the model of interest in one dimension and within a mean field approximation where the correlations between sites are ignored. In Sec.~\ref{sec:jam}, we describe the phase diagram which is obtained using Monte Carlo simulations and mean field calculations in detail. We then develop a perturbation theory for weak activity in Sec.~\ref{pertL} and for strong activity in Sec.~\ref{pertS}. In Sec.~\ref{sec:pn}, the numerical results for the  size distribution of the hole clusters and the distribution of the largest hole cluster are described. We close the article with a summary and some open questions in Sec.~\ref{sec:con}.}

\color{black}
\section{Model}


\subsection{One-dimensional model}
\label{sec:1d}

We consider a system of $N$ hard core particles on a ring with $L$ sites so that the particle density, $\rho=N/L$. Each particle is also endowed with a binary internal variable, $\eta=\pm 1$. The $+(-)$ particle moves to the adjacent site on the right (left), provided it is empty, at rate ${\tilde u}_n$ that depends on the number $n$ of holes in front of it. 
A particle can also reverse  its internal state at a constant rate ${\tilde \gamma} > 0$. An example of these dynamics in time $dt \to 0$ is given below:
 \bea
{\oplus}\Circle\Circle\Circle{\ominus} &\stackrel{{\tilde u}_3 dt}{\longrightarrow}& \Circle{\oplus}\Circle\Circle{\ominus} \\
{\oplus}\Circle\Circle\Circle{\ominus} &\stackrel{{\tilde \gamma} dt}{\longrightarrow}& {\ominus}\Circle\Circle\Circle{\ominus}
\eea
Here, we work with the hop rate, 
\be
{\tilde u}_n=u_\infty \left(1+\frac{b}{n} \right) ~,~b > 2 \label{rate}
\ee
where, $n > 0, u_\infty \equiv {\tilde u}_{n \to \infty}$ and the restriction on the parameter $b$ is explained in Appendix~\ref{app_ep}. Thus, the relevant time scales in our model are $u_\infty^{-1}$ and $\gamma^{-1}$; 
however, for convenience, for the rest of the article, we measure time in units of $u^{-1}_\infty$ and define the dimensionless parameters, 
\be
u_n=\frac{{\tilde u}_n}{u_\infty}, \gamma=\frac{{\tilde \gamma}}{u_\infty} \label{relab}
\ee

By mapping the holes and particles in a configuration, respectively, to particles without hardcore interactions and sites with multi-particle occupancy \cite{Evans:2005}, one obtains a model in which each site also carries the internal variable (sitewise activity). This model differs from those studied in \cite{Thompson:2011,Solon:2015} where a site can support more than one particle but the particles are left- or right-movers (particlewise activity). For $b=0$, the above hop rate is independent of $n$, and our model reduces to a persistent exclusion process \cite{Soto:2014,Slowman:2016,Dandekar:2020,Erignoux:2021} in which the jamming transition does not occur. However, to understand the behavior of the fluid phase for $b > 2$, we briefly study this process. 

{\color{black} As our model does not seem to be exactly solvable, we {investigated it  using} Monte Carlo simulations {in which each step} consists of $L$ update trials. During {each  microstep}, irrespective of its orientation, a particle is chosen uniformly. Then with probability $\frac{\gamma}{1+\gamma}$, the particle  tumbles ({\it i.e.}, reverses the sign of its internal variable $\eta$), otherwise it attempts to hop from its original site $i$ to the neighboring site $i + \eta$, provided it is empty. If there are $n$ holes in the direction of hopping, the particle moves to the chosen site with probability ${\tilde u}_n$ (to ensure ${\tilde u}_n < 1$ for all $n > 0$, $u_\infty < (1+b)^{-1}$ in the simulations). Starting from an initial configuration in which the particles are placed at random on the ring and their {internal state is assigned to be $+1$ or $-1$ with equal probability}, the system is evolved for a time $\sim 100 L^2$ to ensure that it reaches the stationary state, following which the quantities of interest are measured.}

\subsection{Quantities of interest}

In the following, we consider  the probability $P^{\alpha \beta}_n(t)$ of a hole cluster of length $n$ whose left and right boundary, respectively,  has particle of type $\alpha$ and $\beta$ where $\alpha, \beta=\pm 1$ at time $t$. The normalization and particle conservation conditions, respectively, read as
\bea
\sum_{n=0}^N (P^{++}_n(t)+P^{+-}_n(t)+P^{-+}_n(t)+P^{--}_n(t)) &=& \rho \label{norm1}\\
\sum_{n=0}^N n (P^{++}_n(t)+P^{+-}_n(t)+P^{-+}_n(t) +P^{--}_n(t)) &=& 1-\rho  \label{cons1}
\eea
at all times. As a hole cluster of length $n$ is bounded by two particles, the normalization condition \eq{norm1} means that one is summing over hole clusters of all lengths with respect to (say) the right particle with either sign, while the condition \eq{cons1} states that the mean length of the hole cluster is simply the hole density. 

In the following, we are interested in the stationary state where the probability $P^{\alpha \beta}_n(t) \stackrel{t \to \infty}{\to} P^{\alpha \beta}_n$. As we have assumed that the switch between $+$ and $-$ state of a particle occurs at the same rate, the distribution $P^{--}_n=P^{++}_n$. In Secs.~\ref{sec:jam}-\ref{pertS}, we focus on the {\it mean mobility}, $w$ defined as the mean hop rate from a site, given that the site is occupied. {Irrespective of whether a site is occupied by the left- or  right-mover, due to $P^{--}_n=P^{++}_n$, the mean hop rate, $z=\sum_{m=1}^\infty u_m (P^{+-}_m+P^{++}_m)$}. Then we have 
\be
w=\frac{2 z}{\rho} \label{mobidef}
\ee
where, the factor of $2$ ensures that in the limit of vanishing activity, the mobility in the passive model is recovered (see Appendix~\ref{app_ep}). 
In Sec.~\ref{sec:pn}, the probability distributions $P^{++}_n, P^{+-}_n, P^{-+}_n$ are discussed.

\subsection{Mean field theory in grand canonical ensemble}
\label{sec:mft}

In Appendix~\ref{app_dyn}, we describe the time evolution equations for the distributions $P^{\alpha \beta}_n(t)$ within a mean field theory for arbitrary $u_n$ when $N, L \to \infty$ but $\rho$ is finite. For  $\gamma > 0$, 
 the dynamical equations \eq{pm}-\eq{mm} with initial condition $P^{\alpha \beta}_n(0)=\frac{\rho^2}{4} (1-\rho)^n$ and boundary condition $P^{\alpha \beta}_{n_{\max}}(t)=0$ for $\alpha, \beta=\pm 1$ were numerically integrated for $0 \leq n \leq n_{\max}$. The stationary state distributions and the mean mobility were obtained by running these equations for a time $n^2_{\max}$; in all the data shown here, $n_{\max}=1024$. 
 
 On setting the RHS of equations (\ref{pm})-(\ref{mm}) to zero, we obtain the following mean field equations for the steady state probabilities: 
\bea
u_{n+1}P^{+-}_{n+1} - u_{n}P^{+-}_n+ \gamma ( P^{++}_n - P^{+-}_n) &=& 0 \label{Gn2}\\
w (P^{-+}_{n-1} - P^{-+}_n) + \gamma (P^{++}_n - P^{-+}_n) &=&0 \label{Hn2}\\
w (P^{++}_{n-1}- P^{++}_n)  +u_{n+1}P^{++}_{n+1}-  u_n P^{++}_n+\gamma (P^{+-}_n + P^{-+}_n- 2  P^{++}_n) &=& 0 \label{Pn2} 
\eea
with $u_0=P^{-+}_{-1}=P^{++}_{-1}=0$. In Appendix~\ref{app_arbb}, we show that the generating functions corresponding to these distributions obey linear, inhomogeneous, second order differential equations 
with variable coeﬀicients, and the model appears intractable for $b > 0$. However, for $b=0$, explicit expressions for these distributions can be obtained, as discussed in Appendix~\ref{app_pep}.

\section{Transition and crossover in stationary state}
 \label{sec:jam}

\begin{figure}[t!]
     \centering
     \begin{subfigure}{0.49\textwidth}
         \centering
         \includegraphics[width=1.0\textwidth]{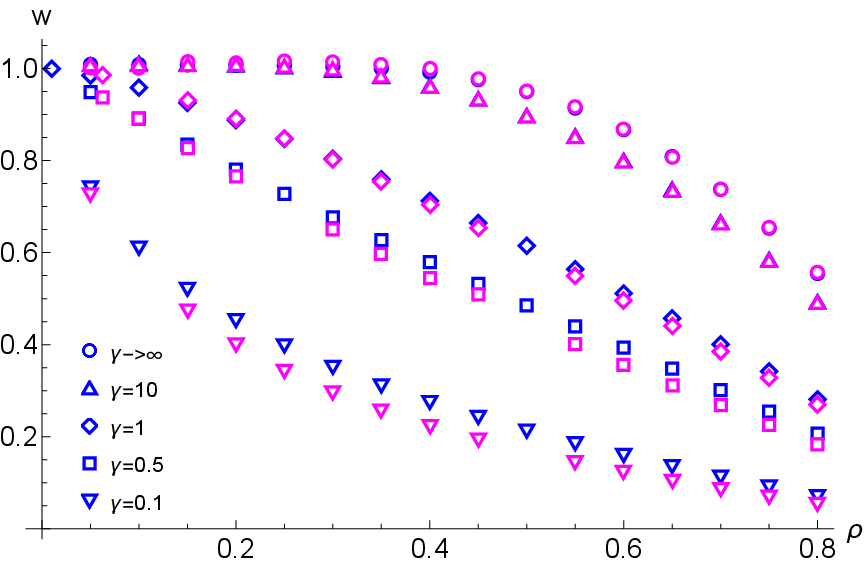}
                 \caption{}
        \label{mobi1}
     \end{subfigure}
     \begin{subfigure}{0.49\textwidth}
         \centering
         \includegraphics[width=1.0\textwidth]{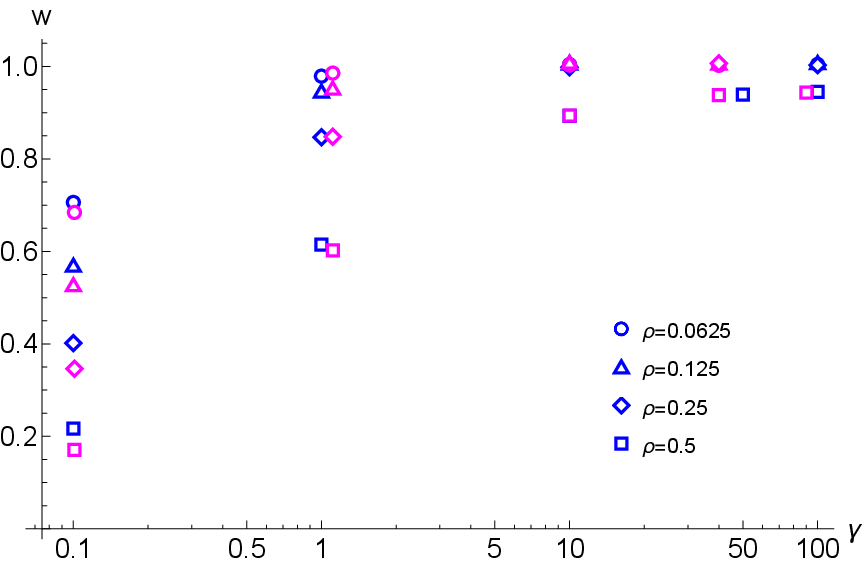}
         \caption{}
        \label{mobi2}
     \end{subfigure}
     \centering
     \begin{subfigure}{0.49\textwidth}
         \centering
         \includegraphics[width=1.0\textwidth]{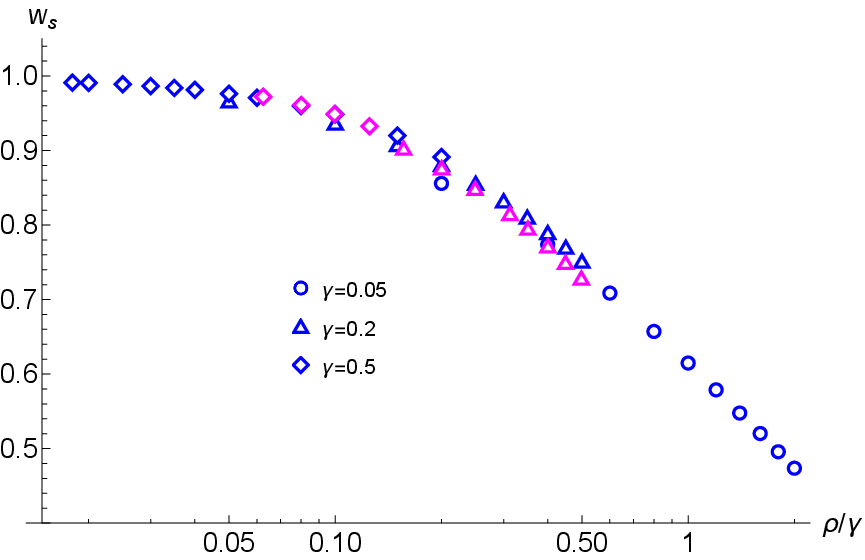}
                 \caption{}
        \label{mobi3}
     \end{subfigure}
     \begin{subfigure}{0.49\textwidth}
         \centering
         \includegraphics[width=1.0\textwidth]{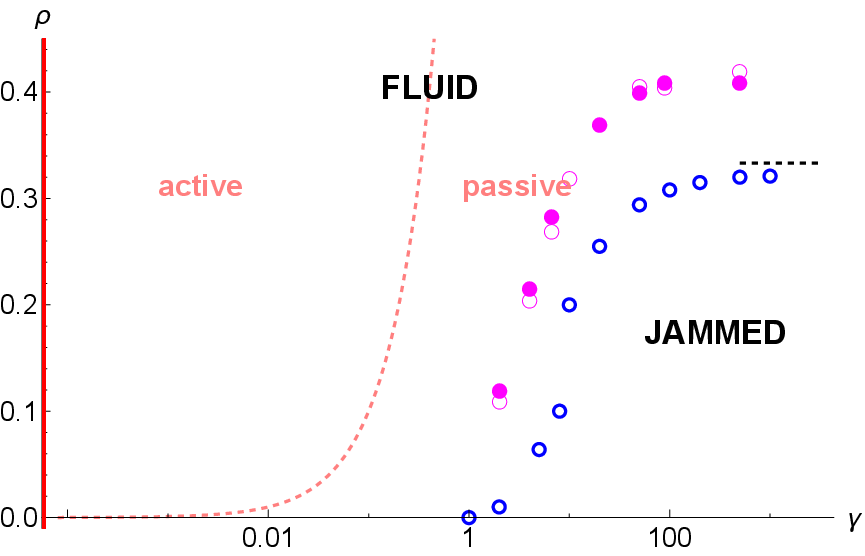}
         \caption{}
        \label{mobi4}
     \end{subfigure}
     \caption{Mean particle mobility defined in \eq{mobidef} for various switching rates and densities with $b=2.5, u_\infty=0.1$. The data are obtained in Monte Carlo simulations of the $1$D model (magenta) with $L=1024$ and by numerically integrating \eq{pm}-\eq{mm} for the mean field model (blue). In (a) and (b), the data from simulations and mean field theory are indistinguishable for large $\gamma$. In (c), the mobility $w_s$ in scaling limits, $\rho, \gamma \to 0$ and finite $\rho/\gamma$ is shown. In (d), the red vertical line denotes the absorbing phase, the orange curve is $\rho=\gamma$ and  the black dashed line corresponds to the critical density ${\cal r}_c=0.33$ in the passive model. {The filled and open magenta points show the simulation data for $L=512$ and $1024$, respectively}.}
\label{mobi}
\end{figure}

When $\gamma \to \infty$, the system is sensitive to the time-averaged $\eta$,  which vanishes as the switching rate between the $+$ and $-$ states is equal. Thus, in this limit, we obtain the passive model for which the stationary state is unique, and the stationary state measure in the one-dimensional and the mean field model is same \cite{Evans:2005}.  As discussed in Appendix~\ref{app_ep}, for $b > 2$, the steady state is characterized by a phase transition at the critical (particle) density ${\cal r}_c$.  For $\rho > {\cal r}_c$, the system is in the {\it fluid phase} where the inter-particle distance is of order one and the mobility ${\cal w}$, determined by \eq{app_cons}, is strictly below one and decreases with increasing particle density. But in the {\it jammed phase} where $\rho < {\cal r}_c$,  a macroscopic hole cluster forms and the mobility is equal to one, refer to Fig.~\ref{mobi1}. Physically, these behavior arise because of the interplay between hardcore interactions that are important at high densities and the hop rate \eq{rate} which (for $b > 0$) decreases with growing hole-cluster length thus promoting jamming at low particle densities. 

On the other hand, when $\gamma=0$, the system eventually reaches an absorbing state in which none of the particles can hop, due to hard core interactions and the quenched internal variable, resulting in zero mean mobility  for any density. These stationary states depend on the initial condition and  are characterized by a mosaic of macroscopic number of particle and hole clusters with the latter bounded by a left (right)-mover on left (right), that is, configurations such as  
$...\textcircled{\textit{p}}{\ominus}\Circle...\Circle{\oplus}\textcircled{\textit{p}}...$, where particle $\textcircled{\textit{p}}$ is of either sign.

Here, we are interested in elucidating the stationary state when $0 < \gamma < \infty$ and the parameter $b > 2$, and  characterize it by the particle mobility $w$. 
Figure~\ref{mobi1} shows that for sufficiently large $\gamma$, the qualitative behavior of the mobility is the same as in the passive model discussed above, but the critical density $\rho_c$ (the maximum density at which the mean mobility is one) decreases with decreasing $\gamma$. However, for  small enough $\gamma$, Fig.~\ref{mobi1} shows that the mobility does not reach one 
and therefore, for sufficiently slow switching, the system remains in the fluid phase for all densities. 
Similarly, as shown in Fig.~\ref{mobi2}, when the particle density is small enough, the mobility increases with $\gamma$ and reaches one at a finite switching rate $\gamma_c$. But for larger densities, the mobility remains below one for any finite $\gamma$. 

Figure~\ref{mobi1} and Fig.~\ref{mobi2} also show that in the fluid phase, the mobility in the active system is smaller than in the passive system. This is a direct consequence of the constraint that an active particle can move to an adjacent hole provided it is correctly oriented; thus the left-mover in a configuration such as $\textcircled{\textit{p}}{\ominus}\Circle$ is jammed and, if all else remains the same, its state must switch to enable it to hop. Such an activity-induced jamming is more likely for small $\gamma$.

To understand the nature of the fluid phase at small switching rates, we first note that if the particle density is small and held fixed, as $\gamma \to 0, w \to 0$ since the system enters an absorbing state when $\gamma=0$, while for fixed small $\gamma$, as $\rho \to 0, w \to 1$ since the hardcore interactions are negligible at low particle densities. Thus, the order in which the $\rho \to 0$ and $\gamma \to 0$ limits are taken matters as the states about which these perturbations occur are different in character in the two limits. Interestingly, as shown in Fig.~\ref{mobi3}, the mobility approaches a nontrivial function, $w_s$ in the scaling limits $\rho, \gamma \to 0$ with $s=\rho/\gamma$ finite. In accordance with the preceding discussion, for $s \to 0+$, the mobility $w_s \to 1-$ 
 and decays to zero as $s \to \infty$. Thus, in the above mentioned scaling regime, the system is in the fluid phase but the mobility $w_s$ shows a crossover from a {\it passive fluid} behavior to an {\it active fluid}, 
 with increasing activity. 

The above discussion is summarized in the phase diagram shown in Fig.~\ref{mobi4}. For $\gamma=0$, the system is in the absorbing phase for any density in the $1$D and mean field model. For $\gamma > 0$, 
 the phase boundary in the simulations and mean field theory, respectively, correspond to the maximum density where the mean mobility is  $0.99$ and $1.003$. The overshoot in the mean mobility is a finite size effect \cite{Chleboun:2010}, {and for this reason, two system sizes were simulated to obtain the phase boundary. For $\gamma$ close to one, to reach our critical density criteria for mobility, very low densities were simulated ($\approx 0.0625$) but, as such a low density data likely suffers from finite size effects, it is not reliable and therefore, it is not shown}. For the parameters in Fig.~\ref{mobi}, our data suggests that  the transition between the fluid and the jammed phase occurs for $\gamma > 1$, but the system stays in the fluid phase for $\gamma < 1$. Within the fluid phase, we demarcate the active fluid and passive fluid behavior by  $\rho=\gamma$ curve, in accordance with \eq{wsc}, which is discussed below.

\section{Perturbation theory for large switching rates}
\label{pertL}

\begin{figure}[t]
     \centering
     \begin{subfigure}{0.49\textwidth}
         \centering
         \includegraphics[width=1.0\textwidth]{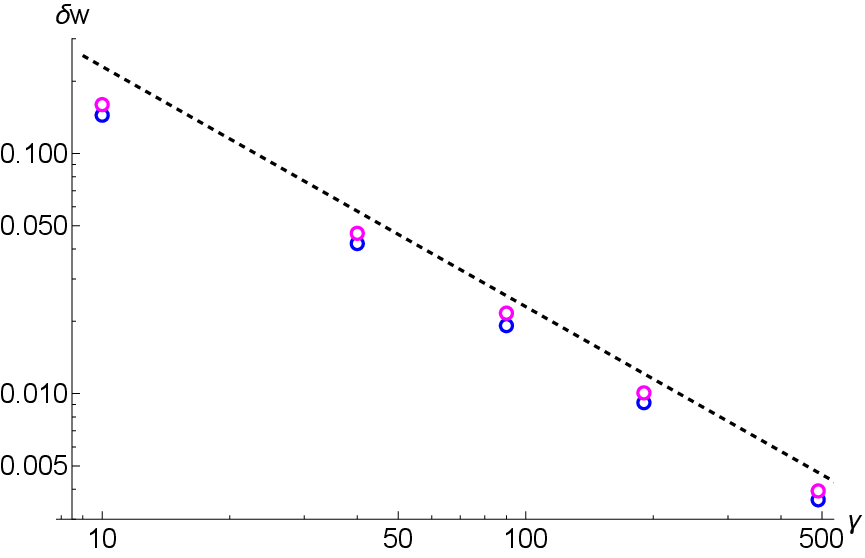}
                 \caption{}
        \label{fug_a}
     \end{subfigure}
     \begin{subfigure}{0.49\textwidth}
         \centering
         \includegraphics[width=1.0\textwidth]{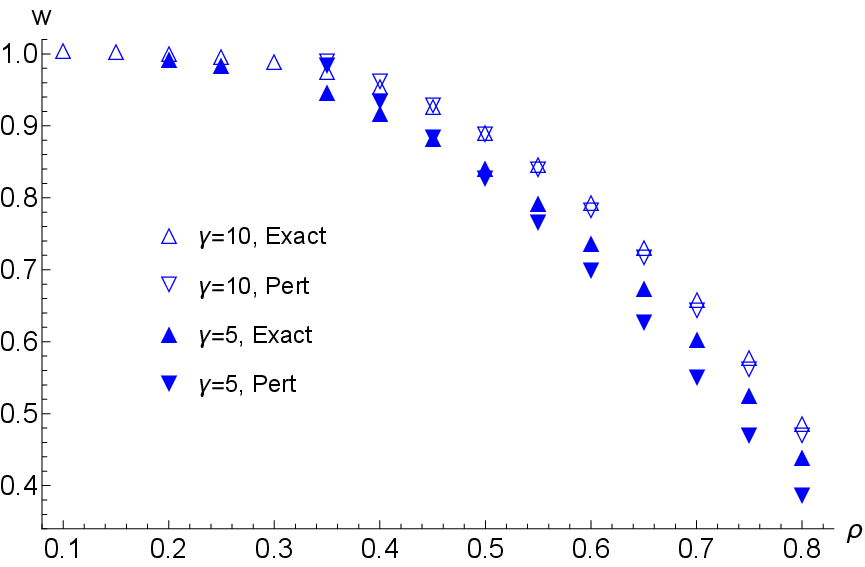}
         \caption{}
        \label{fug_b}
     \end{subfigure}
\caption{(a) Deviation, $\delta w=w(\gamma \to \infty)-w(\gamma)$ in mobility for $b=4.75, \rho={\cal r}_{c}=11/15$ as a function of switching rate for the $1$D model (magenta) and in mean field theory (blue) with the dashed line depicting $\gamma^{-1}$. (b) Mean mobility in the mean field theory as a function of density for $b=2.5$, and $\gamma=10$ (open triangles) and $5$ (filled triangles). The data shown are obtained  by numerically integrating the mean field equations \eq{pm}-\eq{mm} (up-pointing triangles) and using the perturbation theory (down-pointing triangles) described in Sec.~\ref{pertL}.}
\label{fig_fugmft}
\end{figure}

We first consider the parameter regime, $\gamma \gg 1$ (weak activity) and $\rho > {\cal r}_c > \rho_c$ (fluid phase). The key point of the theory outlined below is to show that at a fixed density in the fluid phase, the mobility in the active system decreases with increasing activity. 
Figure \ref{fug_a} shows the deviation in mobility due to weak activity which decays with switching rate as $\gamma^{-1}$. We also note that unlike in Fig.~\ref{mobi1} and Fig.~\ref{mobi2} where the data obtained from simulations and mean field theory appear to coincide for large $\gamma$, Fig.~\ref{fug_a} shows that they do not exactly overlap. 

To capture the deviation in the mobility analytically, we consider the mean field model and 
expand the steady state distributions and mobility in a power series, respectively, as $P^{\alpha \beta}_n = \sum_{k=0}^\infty \gamma^{-k}  p^{\alpha \beta}_{n,k}, \alpha, \beta=\pm 1$ and $w=\sum_{k=0}^\infty \gamma^{-k}  w_k$. Since all the four distributions are identical in the passive limit, 
we have 
$ p^{\alpha \beta}_{n,0}=\frac{{\cal K}_n}{4}$  where, due to \eq{passiveP}, 
\bea
 u_{n+1} {\cal K}_{n+1}-  \left( u_n+{\cal w} \right) {\cal K}_n + {\cal w} {\cal K}_{n-1} &=& 0, n \geq 0 \label{passiveK}
 \eea
 with $w_0={\cal w}$, $u_0={\cal K}_{-1}=0$, and the distribution ${\cal K}_n$ is given by \eq{Pnzrp}.

To find the first order correction to the distributions and mobility due to finite switching rates, using the above expansions in  \eq{Gn2} and \eq{Hn2} and keeping terms to linear order in ${\gamma}^{-1}$, we obtain
\bea
 p^{\pm \mp}_n &=& p^{++}_n \mp \frac{{\cal w}}{4}  ({\cal K}_{n-1} -{\cal K}_n)  \label{Gfeqn}
 \eea
 where we have dropped the subscript $k=1$ for brevity and used \eq{passiveK} to simplify the expressions. On expanding \eq{Pn2} to linear order in ${\gamma}^{-1}$, we obtain $p^{+-}_n+p^{-+}_n=2 p^{++}_n$ which is consistent with \eq{Gfeqn}. Then keeping terms to quadratic order in ${\gamma}^{-1}$ (or alternatively, adding \eq{Gn2}-\eq{Pn2} and expanding the resulting equation to linear orders in ${\gamma}^{-1}$), we obtain
 \bea
 &&u_{n+1} p^{++}_{n+1}-  \left( u_n+{\cal w} \right) p^{++}_n +{\cal w} p^{++}_{n-1}+2 w_1 ({\cal K}_{n-1}-{\cal K}_n) \nn \\
 &+& u_{n+1} p^{+-}_{n+1}- u_n p^{+-}_n +{\cal w} (p^{-+}_{n-1}-p^{-+}_n) =0
 \eea
Due to \eq{Gfeqn}, the above equation simplifies to give 
\bea
u_{n+1} p^{++}_{n+1}-  \left( u_n+{\cal w}\right) p^{++}_n + {\cal w} p^{++}_{n-1}=q_n, n \geq 0 \label{Pinhomo}
\eea
where, $p^{++}_{-1}=0$ and $q_n$ is given by \eq{app_qn}. 
Thus, we find that the first order correction, $p^{++}_n$ obeys an inhomogeneous equation with the homogeneous equation being the same as that for ${\cal K}_n$. 

As detailed in Appendix~\ref{app_pertL}, on solving \eq{Pinhomo}  and invoking the constraints and the boundary conditions, the deviation in mobility can be obtained.  
It should be noted that the analyses in this section is valid only when the density is in the fluid phase of the passive
model, that is, $\rho > {\cal r}_c$  (see Appendix~\ref{app_pertL} for an explanation). 
For $b=0$, from the above analyses, we find that $w_1 =-\frac{\rho}{2} (2-\rho) (1-\rho)$ which matches that obtained from the exact solution (see \eq{wb0larg}). 
For $b \neq 0$, although an explicit expression for $w_1$ can be obtained, as it is very long, we do not display it here; however, in Fig.~\ref{fug_b}, the mobility obtained using the above perturbation theory is compared against that by numerically integrating the mean field equations for $\rho > {\cal r}_c$; we find that the above analyses is quite accurate for $\gamma=10$ but significant deviations occur for smaller $\gamma$. 

\section{Perturbation theory for small switching rates}
\label{pertS}

We now consider the parameter regime, $\rho, \gamma \to 0$ with $s=\rho/\gamma$ large but finite and $b \geq 0$, and obtain an expression for the mobility in the mean field model. As already discussed in Sec.~\ref{sec:jam}, when the system is at a finite, small density and has $\gamma=0$, it eventually enters an absorbing state where the mobility is zero and the hole cluster of length $n > 0$ is only of ${\ominus}\Circle...\Circle{\oplus}$ type, that is, $P_n^{-+} \neq 0, n \geq 0$ while $P_n^{+-}, P_n^{++} \propto \delta_{n,0}$.  

If now $\gamma$ takes a small nonzero value, 
 one would obtain nonzero $P_n^{++}, n > 0$ if the sign of the particle on either edge of the ${\ominus}\Circle...\Circle{\oplus}$ cluster flips, and therefore, we expect  $P_n^{++} \sim O(\gamma), n > 0$; in a similar manner, it can be argued that  the distribution $P_n^{+-} \sim O(\gamma^2), n > 0$. 
As a result, in \eq{Gn2}, the last term on the LHS can be neglected for $n > 0$, but for $n=0$, 
all the terms are required in order to avoid negative or vanishing probabilities. These considerations then yield 
\bea
u_1 P^{+-}_{1}+ \gamma ( P^{++}_0 - P^{+-}_0) &=& 0 \label{G0sca}\\
u_{n+1}P^{+-}_{n+1} - u_{n}P^{+-}_n+ \gamma P^{++}_n &\approx& 0, n > 0 \label{Gnsca}
\eea
From \eq{G0sca}, it also follows that $P^{++}_0, P^{+-}_0 \sim O(\gamma)$. 

We next consider \eq{Hn2}: as the mobility is finite and $P_n^{++} \sim O(\gamma), n \ge 0$, 
the last term on the LHS of  \eq{Hn2} can be neglected, and we  obtain 
\bea
w (P^{-+}_{n-1} - P^{-+}_n) + \gamma P^{++}_n  &\approx&0, n \geq 0 \label{Hnsca}
\eea
which suggests that $P^{-+}_n \sim O(\gamma^2)$. At first sight, this $\gamma$-dependence seems problematic as the distribution $P^{-+}_n$ is nonzero for $\gamma=0$. Then, for consistency, we expect that $P^{-+}_n \propto \gamma^2 s^2$, which is indeed found from the exact solution for $b=0$ in Appendix~\ref{app_pep}. Finally, from \eq{Pn2}, we can write
\bea
w (P^{++}_{n-1}- P^{++}_n)  +u_{n+1}P^{++}_{n+1}-  u_n P^{++}_n&\approx& 0 , n \geq 0 \label{Pnsca}
\eea
Equations \eq{G0sca}-\eq{Pnsca} are analyzed in Appendix~\ref{app_pertS}, and we find that  the mobility, $w=\gamma/\epsilon$ where $\epsilon$ is a solution of \eq{eptrns} which can be approximated to finally yield $w \stackrel{s \gg 1}{\approx} 4/s, b \geq 0$. 

\begin{figure}[t]
     \centering
     \begin{subfigure}{0.49\textwidth}
         \centering
         \includegraphics[width=1.0\textwidth]{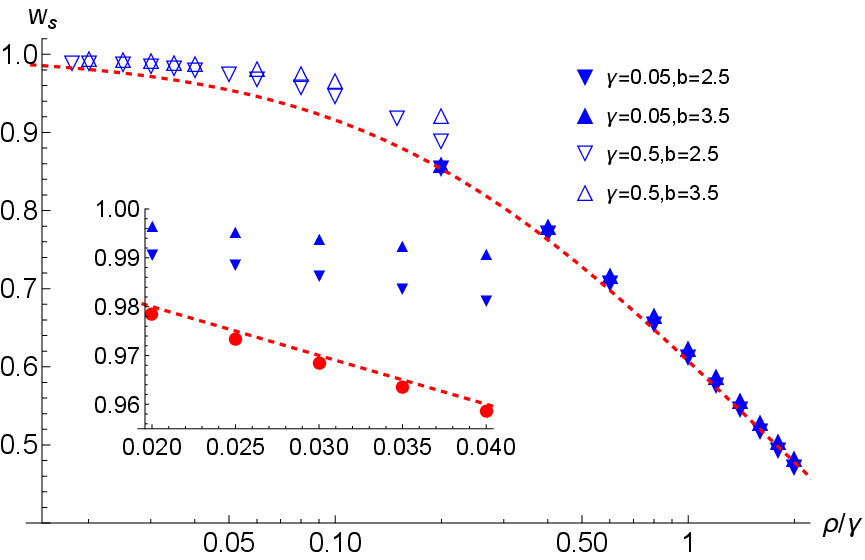}
                 \caption{}
        \label{mobib1}
     \end{subfigure}
     \begin{subfigure}{0.49\textwidth}
         \centering
         \includegraphics[width=1.0\textwidth]{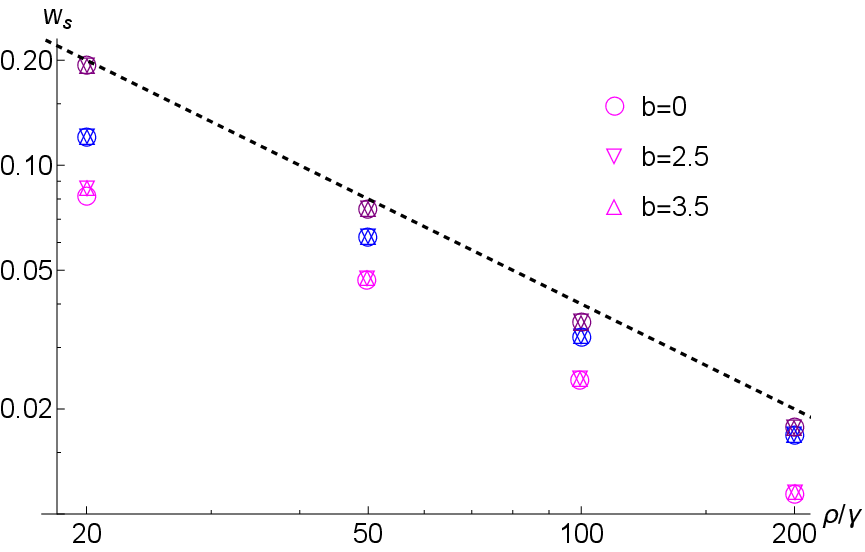}
         \caption{}
        \label{mobib2}
     \end{subfigure}
     \caption{Mean mobility $w_s$ in the scaling limits, $\rho, \gamma \to 0$ with finite $\rho/\gamma$. 
     (a) The points are obtained by numerically integrating the mean field equations \eq{pm}-\eq{mm} for $b=0$ (circles), and $b=2.5$ and $3.5$ (triangles). The dashed line shows the mean mobility for $b=0$ given by \eq{wscube} (main figure) and \eq{wb0s} (inset). 
     (b) The points are obtained from simulations (magenta), numerically integrating \eq{pm}-\eq{mm} (blue) and solving \eq{eptrns} (purple). The dashed line  shows the curve $4/x$.}
\label{fig_mobscb}
\end{figure}

In Fig.~\ref{fig_mobscb}, the mean mobility obtained by numerically integrating \eq{pm}-\eq{mm} and numerically solving \eq{eptrns} is shown in the scaling regime for various parameters. Figure~\ref{mobib1} shows that the mean mobility for $b > 0$ is not significantly different from that for $b=0$, and 
that the mobility decreases linearly with $s$ when $s \ll 1$. 
In Fig.~\ref{mobib2}, the mobility for $s \gg 1$ is shown, and we find that the results obtained from  simulations, exact mean field equations, approximate equations \eq{G0sca}-\eq{Pnsca} and the analytical approximation $4/s$ agree for large $s$. 
The numerical and analytical results described above can thus be summarized as 
\bsn
{w_s \approx \label{wsc}}
1- \frac{B(b) \rho}{\gamma} &,  $\rho \ll \gamma$ \\
\frac{4 \gamma}{\rho} &, $\rho \gg \gamma$
\esn
where $B(b) > 0$ is a decreasing function of $b$ with $B(0)=1$, due to \eq{wb0s}. In Sec.~\ref{sec:jam}, these behavior corresponding to large and small switching rate are, respectively, termed passive fluid and active fluid, and demarcated by $\rho=\gamma$ curve in Fig.~\ref{mobi4}.

\section{Size distribution of hole clusters} 
\label{sec:pn}

In the preceding sections, we have focused on a summary statistic, namely, the mean mobility. We now describe the numerical results for the probability distribution of the hole clusters in the stationary state. 


\subsection{Distribution of hole clusters}

{\color{black} 
In Fig.~\ref{fig:dist}, for a given $b$ and $\rho < {\cal r}_c$, the hole cluster distributions are shown for different values of  $\gamma$.  As discussed in Appendix~\ref{app_ep}, in the jammed phase, the hole cluster distribution consists of a power law distribution which is independent of system size and a piece that scales linearly with $L$, while the distribution decays exponentially with an algebraically decaying prefactor in the fluid phase. For $\gamma \gg 1$ where $P_n^{\alpha \beta} \approx \frac{{\cal K}_n}{4}$ for $\alpha, \beta=\pm 1$ (refer to Sec.~\ref{pertL}), the distributions shown in Fig.~\ref{dist1} and Fig.~\ref{dist2} are consistent with these expectations.  But for $\gamma \lesssim 1$, the three distributions are distinctly different, and more importantly, the distribution is now exponential (note the change in $x$-scale in Fig.~\ref{dist3} and Fig.~\ref{dist4}).  

\begin{figure}[t!]
\begin{subfigure}{0.49\textwidth} 
\centering 
\includegraphics[scale=0.65]{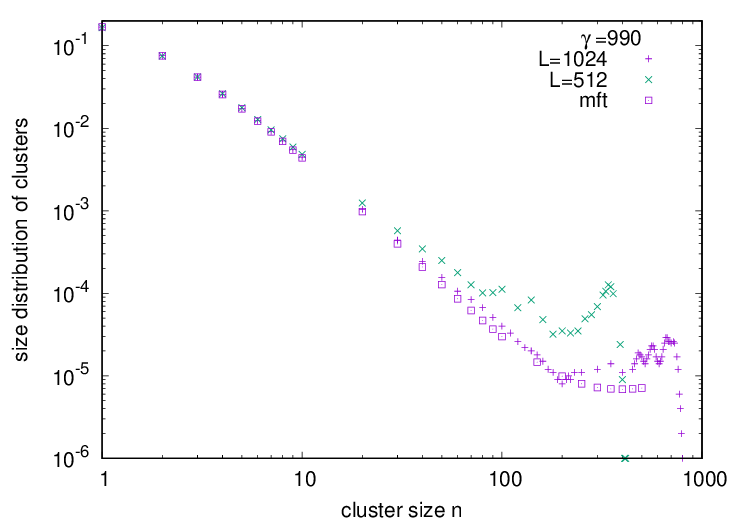} 
\caption{} 
\label{dist1}
\end{subfigure}
\begin{subfigure}{0.49\textwidth} 
\centering 
\includegraphics[scale=0.65]{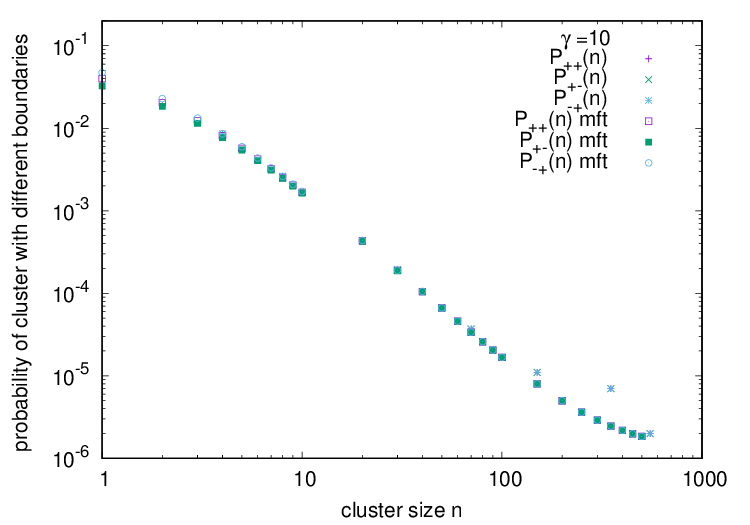} 
\caption{}
\label{dist2}
\end{subfigure}
\begin{subfigure}{0.49\textwidth} 
\centering 
\includegraphics[scale=0.65]{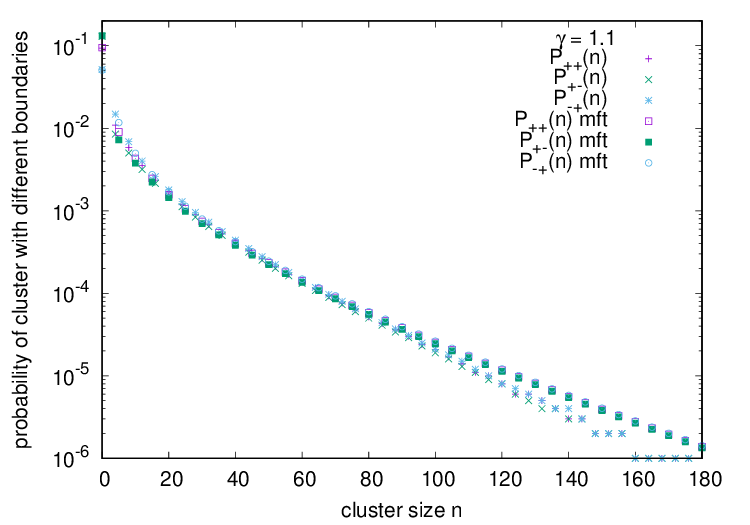} 
\caption{}
\label{dist3}
\end{subfigure}
\begin{subfigure}{0.49\textwidth} 
\centering 
\includegraphics[scale=0.65]{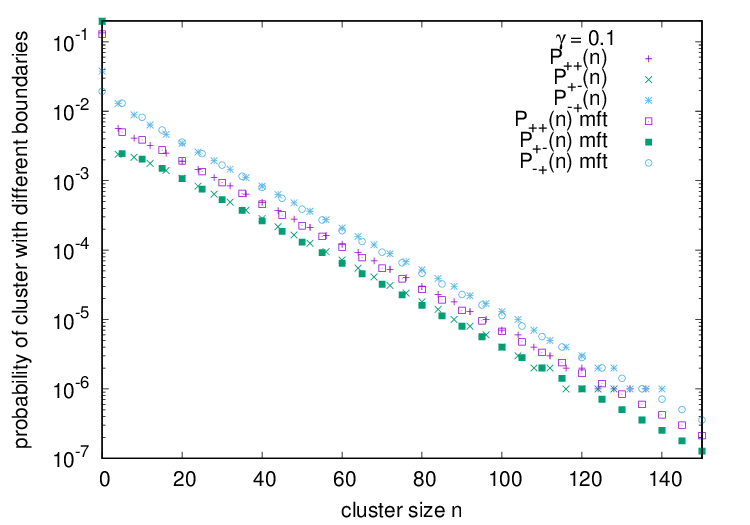} 
\caption{}
\label{dist4}
\end{subfigure}
\caption{{\color{black} Size distribution of hole clusters for $b = 2.5$ and $\rho= 0.125 < {\cal r}_c=0.33$ and various switching rates. The data are obtained from simulations and mean field theory (mft). In (a), $P(n)=\sum_{\alpha, \beta=\pm 1} P_n^{\alpha \beta}$ is shown, while in (b)-(d), the different cluster types are shown separately. }}
\label{fig:dist}
\end{figure}


\subsection{Statistics of the largest hole cluster}
\label{sec:hmax}

{\color{black} 

In Fig.~\ref{max1} and Fig.~\ref{max2}, the distribution of the largest hole cluster (denoted by $h_{\max}$) obtained by averaging over the particle orientations is shown for a density $\rho=0.4$ for which a macroscopic hole cluster in the passive limit is formed for $b=3.5$, but not for $b=2.5$. The data for $b=2.5$ shows that the distribution is nonmonotonic and decays exponentially for all the switching rates. But for $b=3.5$, the largest hole cluster distribution changes qualitatively as $\gamma$ increases: for small $\gamma$, the distribution decays exponentially, but for large $\gamma$ where the system shows jamming, the decay is not exponential,  the distribution becomes wider and the peak position occurs at large cluster sizes.

\begin{figure}[t]
\begin{subfigure}{0.49\textwidth}
\centering
    \includegraphics[scale=0.6]{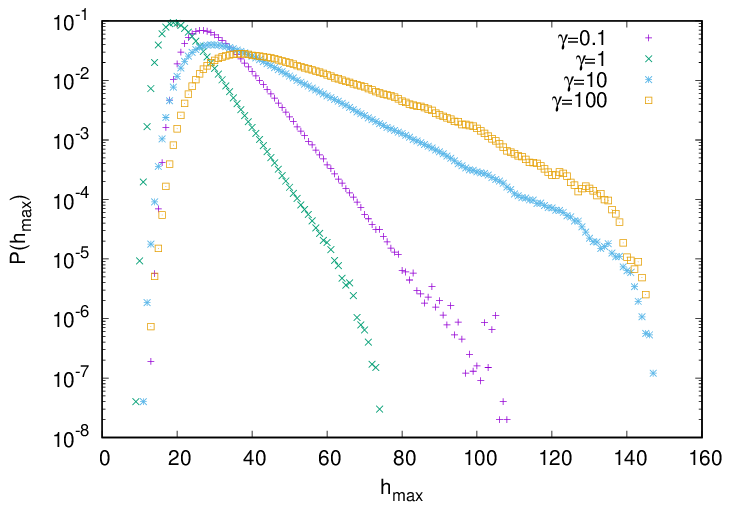} 
    \caption{}
    \label{max1}
\end{subfigure}
\begin{subfigure}{0.49\textwidth}
    \includegraphics[scale=0.6]{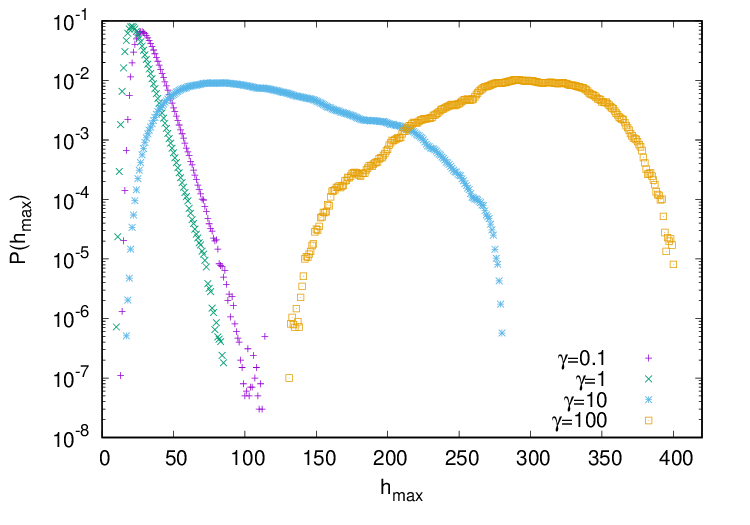} 
    \caption{}
    \label{max2}
  \end{subfigure}  
  \begin{subfigure}{0.49\textwidth}
    \includegraphics[scale=0.34]{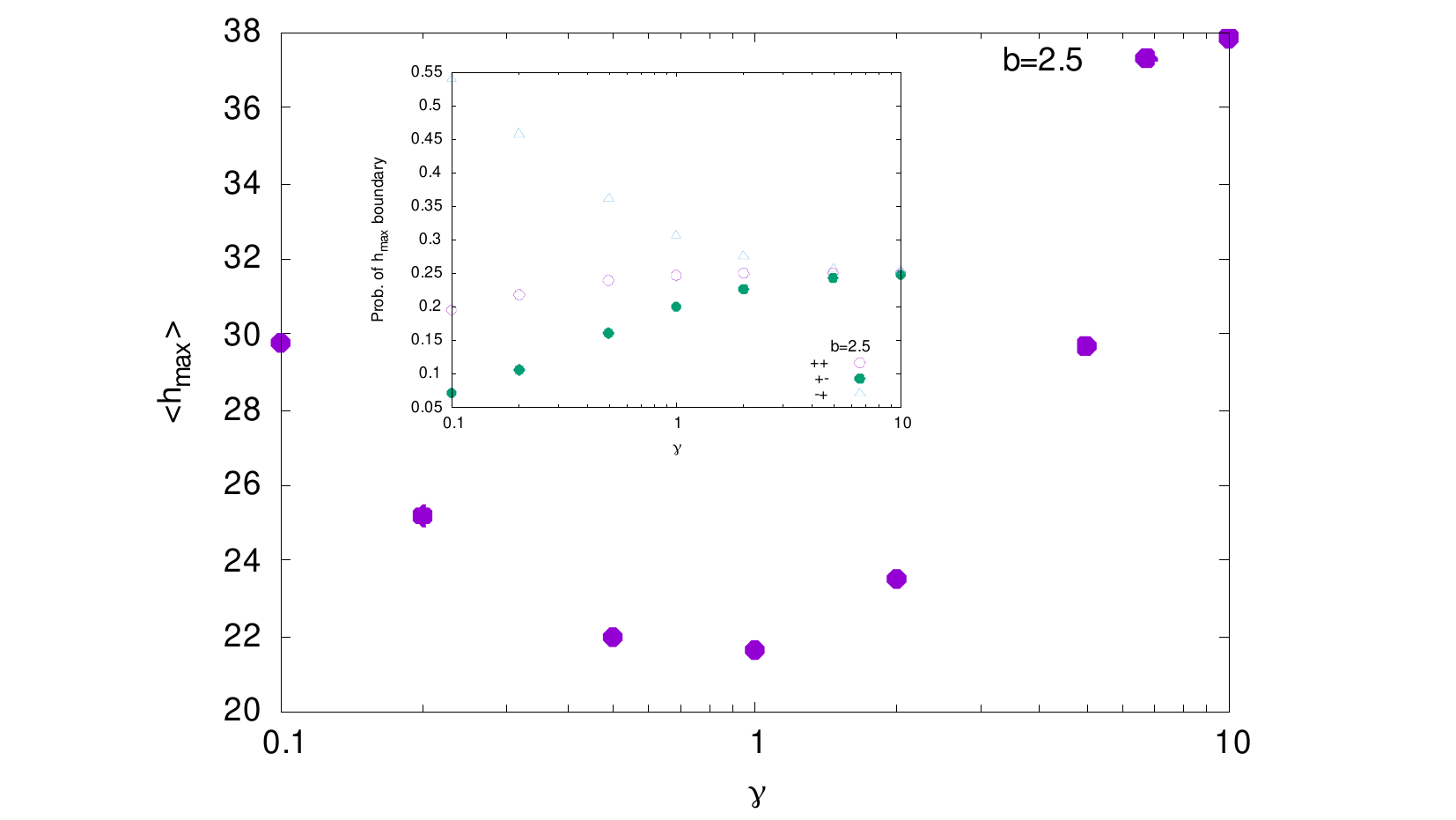} 
    \caption{}
    \label{max3}
  \end{subfigure}  
  \begin{subfigure}{0.49\textwidth}
    \includegraphics[scale=0.34]{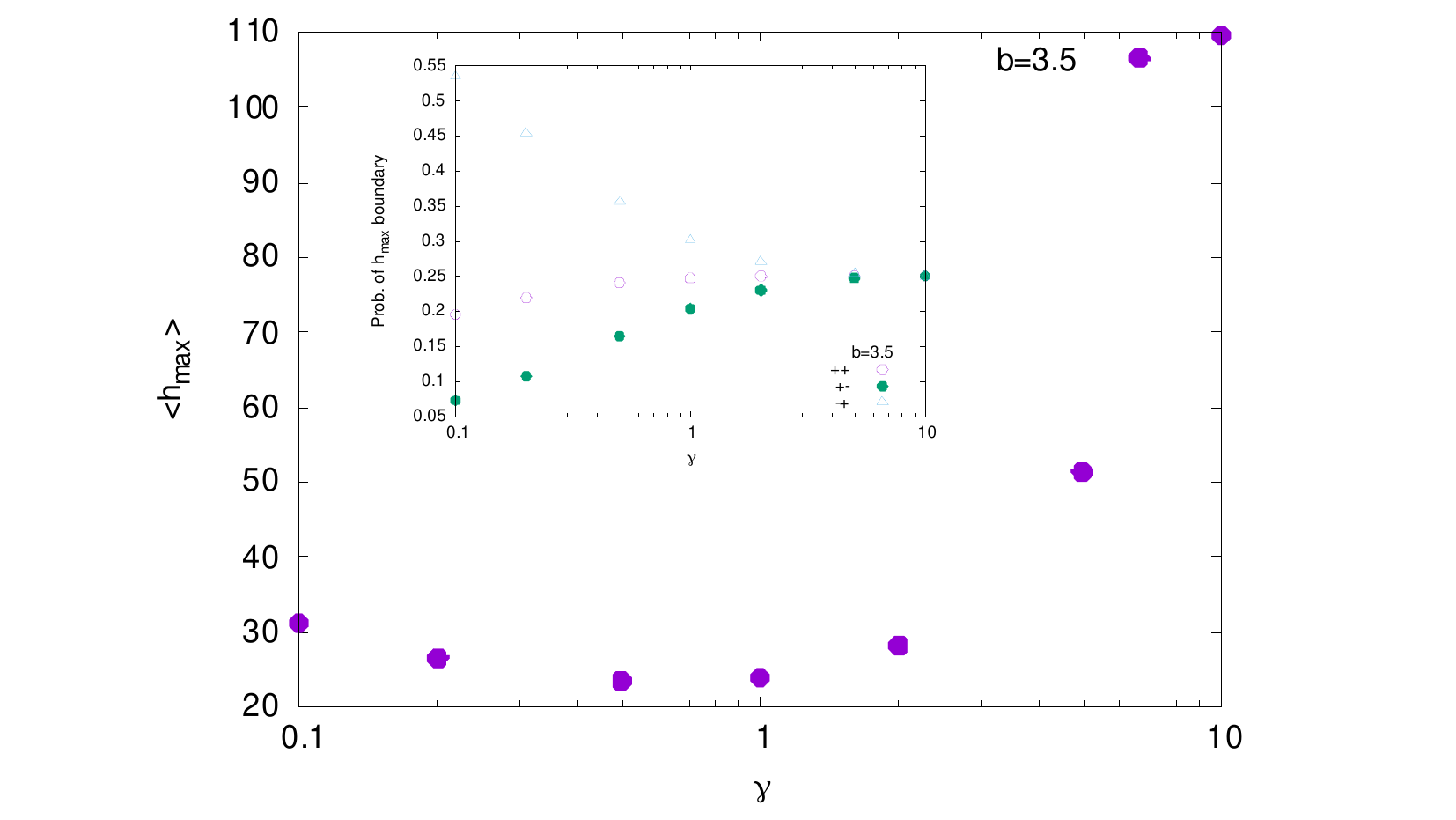}
    \caption{}
    \label{max4}
  \end{subfigure}  
    \caption{Probability distribution of the largest hole cluster size for various values of $\gamma$ with (a) $b=2.5$ and (b) $b=3.5$. (c) and (d): The main figure shows the average size of the largest cluster as a function of $\gamma$. The inset shows the probability of finding the largest hole cluster of a specific type as a function of $\gamma$.  In all the plots, $L=1024$, $\rho=0.4$, $u_\infty = 0.1$.}
    \label{fig:phmax}
\end{figure}

\color{black}

In Fig.~\ref{max3} and Fig.~\ref{max4}, the average size $\langle h_{max} \rangle$ of the largest hole cluster as a function of $\gamma$ is shown, and for both $b=2.5$ and $3.5$, it exhibits a minimum  when the system is in the fluid phase. This interesting effect can be explained as a result of competition between the switching  and hopping, as described below. The inset of Fig.~\ref{max3} and Fig.~\ref{max4} show the probability that the largest hole cluster is bounded by particle of type $\alpha (\beta)$ on the left (right). For small $\gamma$, as the $-+$ type has the highest probability, it contributes most to $\langle h_{max} \rangle$ so we now focus on this type. 

A hole cluster of $-+$ type can only expand or hold its current size, but can not shrink. For small enough $\gamma$, when the switching rate is much less than the hopping rate ($\gamma \ll 1$), the hole cluster keeps expanding until both boundaries get stuck. But as $\gamma$ increases, it is possible that before a given cluster reaches its maximum possible size (limited only by boundary particles getting stuck), a tumble occurs at one of the boundaries and the expansion stops. This decreases $\langle h_{max} \rangle$ for small $\gamma$. On the other hand, when $\gamma$ is large, as the inset of Fig.~\ref{max3} and Fig.~\ref{max4} show,  all four types of boundaries for the largest hole cluster are equally likely. Now, if a large hole cluster is shrinking from one end, the shrinking continues until the active particle at that end tumbles. As $\gamma$ decreases, the shrinking therefore continues for longer duration. On the other hand, if a hole cluster is expanding at one end, the expansion may stop either because of tumbling, or because of the particle at the boundary getting blocked by another oppositely moving particle. For smaller $\gamma$, the second scenario becomes more likely and therefore the expansion is halted earlier for smaller $\gamma$. This is the reason behind decreasing $\langle h_{max} \rangle$ with decreasing $\gamma$ for large $\gamma$ values.}

\color{black}

\section{Discussion}
\label{sec:con}

In this article, we studied the stationary state of a one-dimensional model of active hard core particles with conserved density $\rho$. In our model, a particle is either a left- or right-mover, and its direction of self-propulsion can switch at rate $\gamma$. A particle can hop to an empty neighbor at rate given by \eq{rate}, provided the particle is oriented in the direction of the target site. 
Using Monte Carlo simulations and a mean field theory, we find that the jamming transition occurs in our model provided the activity is sufficiently weak (large $\gamma$) and for strong enough activity, the system exists only in the fluid phase; the phase diagram in the $\gamma$-$\rho$ plane is summarized in Fig.~\ref{mobi4}. We remark that here we have worked with values of parameter $b$ for which a jamming transition occurs in the passive model. 
But,  for a given $b$, as the critical density $\rho_c$ decreases with decreasing $\gamma$ (refer to Fig.~\ref{mobi4}), we expect that for a given $\rho$, the critical $b$ increases.  

In much of the article, we have compared the results for the $1$D model with a mean field theory. 
As the stationary state measure in the passive model is independent of the dimensionality of the system \cite{Evans:2005}, on extrapolating this result to large $\gamma$, we expect a good match between the two models. For weak activity, our data is in agreement with this expectation and for strong activity, although the results from the two models do not quite overlap, the deviations are not found to be significantly large. 
In Sec.~\ref{pertS}, we developed a scaling theory in the low density, high activity regime ($\rho, \gamma \to 0$ with $s=\frac{\rho}{\gamma}$ finite) which, to our knowledge, is novel. We find that the mobility vanishes in a {\it universal} fashion; more precisely, the mobility decays as $4/s, s \gg 1$ for all $b \geq 0$. 
 
 Our analytical calculations mentioned above are, however, limited to the fluid phase, and a detailed understanding of the effect of activity in the jammed phase is desirable. Although the results obtained from simulations and numerical solution of the mean field theory strongly suggest that, as in the passive limit, the mobility is bounded above by one in the jammed phase, unlike for the passive model \cite{Evans:2005}, we have not shown this result rigorously and it would be useful to place this on a stronger footing. 
Here, we have focused only on the stationary state, but it would be interesting to  investigate the dynamics in the steady state and the coarsening dynamics in the jammed phase.

\clearpage
\appendix
\makeatletter
\renewcommand{\@seccntformat}[1]{Appendix \csname the#1\endcsname\quad}
\makeatother
\renewcommand{\thesection}{\Alph{section}}
\numberwithin{equation}{section}

\color{black}
\section{Passive exclusion process with hole-dependent rates}
\label{app_ep}

For a self-contained discussion, here we review the passive symmetric exclusion process with hop-dependent rates given by \eq{rate}  in which the particles do not carry the internal variable. If ${\cal K}_n(t)$ denotes the probability of a $n-$hole cluster at time $t$ in an infinitely large system with finite density $\rho$, we have
\bea
\partial_t {\cal K}_n(t) &=& 2 {u}_{n+1} {\cal K}_{n+1}(t)-  2\left( {  u}_n+\frac{{\mathcal z}}{\rho} \right) {\cal K}_n(t) + \frac{2 {\mathcal z}}{\rho} {\cal K}_{n-1}(t), n \geq 0 \label{passiveP}
\eea
where $u_n$ is given by \eq{relab}, $u_0={\cal K}_{-1}(t)=0$, and the mean hop rate ${\mathcal z}(t)=\sum_{n=1}^\infty {u}_n {\cal K}_n(t)$ (for an explanation of the factor $2/\rho$, see Appendix~\ref{app_dyn}).

In the stationary state where the LHS is zero, on using the boundary conditions in the equation for $n=0$ gives $u_{1} {\cal K}_{1}=\frac{{\cal z}}{\rho} {\cal K}_0$; repeating this procedure for $n > 0$, we obtain  $u_{n+1} {\cal K}_{n+1}=\frac{{\cal z}}{\rho} {\cal K}_n, n \geq 0$ which can be easily iterated to give
\bea
{\cal K}_n &=& \left(\frac{{\cal z}}{\rho} \right)^n \frac{{\cal K}_0}{\prod_{i=1}^n u(i)} =\left(\frac{{\cal z}}{\rho} \right)^n \frac{{\cal K}_0}{\prod_{i=1}^n (1+\frac{b}{i})}\label{Pnzrp}
 \eea
 Using the normalization condition, $\sum_{n=0}^\infty {\cal K}_n=\rho$, we find that ${\cal K}_0={\rho}/{g(\frac{{\cal z}}{\rho})}$ 
where
\be
g(v)=\sum_{n=0} v^n \frac{n!}{(b+1) (b+2) ... (b+n)} =\, _2F_1(1,1;b+1;v) \label{gdefn}
\ee
and $\, _2F_1(1,1;b+1;v)$ is the Gauss hypergeometric function. 
The mean hop rate (or fugacity; see \cite{Evans:2005}) $\cal z$ is determined from the hole conservation condition, $\sum_{n=0}^\infty n {\cal K}_n=1-\rho$ so that 
\bea
\frac{v g'(v)}{g(v)}\bigg|_{v=\frac{{\cal z}}{\rho}} &=& \frac{\frac{{\cal z}}{\rho} \, _2F_1(2,2;b+2;\frac{{\cal z}}{ \rho})}{(b+1) \, _2F_1(1,1;b+1;\frac{{\cal z}}{ \rho})}=\frac{1-\rho}{\rho} \label{app_cons}
 \eea
For $b=0$ in \eq{rate}, the above equation simplifies to yield the mobility, 
\be
{\cal w} \equiv \frac{{\cal z}}{\rho}=1-\rho  \label{p0fug}
\ee

For $b > 2$, the sum $g({\cal w})$ converges for ${\cal w} \leq 1$. Using $\frac{v g'(v)}{g(v)}\bigg|_{{\cal w}=1} =\frac{1}{b-2}$ in \eq{app_cons}, we obtain the passive critical density to be 
 \be
 {\cal r}_{c} =\frac{b-2}{b-1}~,~b > 2 \label{critpas}
 \ee
 For $\rho > {\cal r}_c$ where ${\cal w} < 1$, as \eq{Pnzrp} shows, the hole cluster length is exponentially distributed (fluid phase), while for $\rho={\cal r}_c$,  the hole cluster size distribution decays algebraically, ${\cal K}_n \sim n^{-b}$ and the variance of the hole cluster size diverges for $2 < b \leq 3$. For $\rho < {\cal r}_c$, in addition to hole clusters whose size varies according to the distribution \eq{Pnzrp} at the critical density, a hole cluster with macroscopic mean length  also exists. But the distribution of this cluster is not captured by the above analysis as it is valid only for a thermodynamically large system.

 For a finite system \cite{Evans:2006}, the exact distribution of a configuration can be obtained by mapping it to a Zero Range Process (ZRP) on a ring with $N$ sites and $M=L-N$ particles. In the ZRP, a site can be occupied by any number of particles and a particle hops to either neighbor at a rate $u_n$ where $n > 0$ is the number of particles at the departure site. Then the steady state measure is a product measure with the constraint that the total mass $M$ is conserved (this statement, in fact, holds in any dimension) \cite{Evans:2005}.  Treating the site and particles in the ZRP, respectively, as hard-core particle and holes in front of the particle, one obtains the exclusion process with hole-dependent rates.

\section{Dynamical equations in mean field theory}
\label{app_dyn}

In the mean field model, the  hole cluster distributions, $P_n^{\alpha \beta}(t), \alpha, \beta=\pm 1$ obey coupled and nonlinear equations as shown below. In the following discussion, for brevity, we will refer to a hole cluster bounded by particle of type $\alpha$ and $\beta$ as $\alpha \beta$ cluster. 
First, consider the distribution $P^{+-}_n(t)$ which obeys the following equation:
\bea
 \partial_t P^{+-}_n(t) &=& 2u_{n+1}P^{+-}_{n+1}(t) - 2u_{n}P^{+-}_n(t) + \gamma [P^{--}_n(t) +  P^{++}_n(t) - 2  P^{+-}_n(t)] \label{pm}
\eea
The first term on the RHS of \eq{pm} arises because a $+-$ hole cluster of size $n$ 
can be created if the particle on either end of a $+-$ cluster with $n+1$ holes hops with rate $u_{n+1}$ in the respective direction; similarly, a $n$-hole $+-$  cluster is destroyed if either $+$ or $-$ particle at the end of the cluster hops in the cluster thus yielding the second term. The next three terms with the coefficient $\gamma$ arise because the sign of the particle type can flip. Due to hard-core interaction between particles, the second term in \eq{pm} is absent when $n=0$, or, in other words, $u_0=0$. 

Next consider the time evolution of $P^{-+}_n(t)$ for which we have
\bea
\partial_t P^{-+}_n(t) &=& \frac{2}{\rho} \{z^{+-}(t)+z^{--}(t)\}\{P^{-+}_{n-1}(t) -P^{-+}_n(t)\}  + \frac{2}{\rho} \{P^{-+}_{n-1}(t) - P^{-+}_n(t) \}  \{z^{+-}(t)+z^{++}(t) \} \nn \\
&+& \gamma [P^{--}_n(t) + P^{++}_n(t) - 2 P^{-+}_n(t)] \label{mp} 
\eea
where, $z^{\alpha \beta}(t)=\sum_{m=1}^\infty u_m P^{\alpha \beta}_m(t)$ is the mean hop rate of $\alpha \beta$ cluster. In one dimension, a $n$-hole $-+$ cluster can be created if there is a $+-$ cluster of length $m > 0$ on the left of a $-+$ cluster with $n-1$ holes as the $-$ particle at the common boundary of these clusters can hop  left with rate $u_m$. In the mean field approximation, the joint distribution factorizes to $P^{+-}_m(t)  P^{-+}_{n-1}(t)$, but then the boundary $-$ particle  is overcounted, and the hop rate is therefore $u_m$ divided by $2/\rho$ as the {\color{black} mean field} probability that the common site is occupied by $-$ particle is $\rho/2$.  Summing over all $m > 0$, we obtain the gain term $\frac{2}{\rho} z^{+-}(t) P^{-+}_{n-1}(t)$ on the RHS of the above equation, and in a similar manner, the other terms in the first line on the RHS of \eq{mp} can be obtained. The last three terms on the RHS of \eq{mp} are due to activity, as in the above equations for $P^{+-}_n(t)$. The above equation holds for $n=0$ also with boundary conditions $P^{-+}_{-1}(t)=P^{++}_{-1}(t)=0$. 

Similarly, one can write the following evolution equations, 
\bea
\partial_t P^{++}_n(t) &=& \frac{2}{\rho} \{P^{++}_{n-1}(t) - P^{++}_n(t)\}  \{z^{+-}(t) +z^{++}(t) \} +u_{n+1}P^{++}_{n+1}(t)-  u_n P^{++}_n(t)
\nn \\
&+&  \gamma [P^{+-}_n(t) + P^{-+}_n(t) - 2  P^{++}_n(t)] \label{pp} \\
\partial_t P^{--}_n(t) &=& \frac{2}{\rho}\{ z^{+-}(t) +z^{--}(t) \}  \{P^{--}_{n-1}(t) - P^{--}_n(t)\}+u_{n+1}P^{--}_{n+1}(t) - u_n P^{--}_n(t)  \nn \\
&+& \gamma [P^{-+}_n(t) + P^{+-}_n(t) - 2 P^{--}_n(t)] \label{mm}
\eea
with $P^{++}_{-1}(t)=P^{--}_{-1}(t)=0$.

\section{Generating function for arbitrary $b$}
\label{app_arbb}

Consider the following generating functions for the steady state distributions that obey the mean field equations \eq{Gn2}-\eq{Pn2},  
 \bea
{\cal G}(v)= \sum_{n=0}^\infty P_n^{+-} v^n , {\cal H}(v)= \sum_{n=0}^\infty  P^{-+}_n v^n, {\cal P}(v)= \sum_{n=0}^\infty  P^{++}_n v^n \label{genfns}
\eea
 We first consider \eq{Hn2} which is independent of the hop rates, and  immediately obtain
\bea
(v-1) w {\cal H}+\gamma ({\cal P}-{\cal H})=0 \label{calHg}
\eea
Similarly, \eq{Gn2} and \eq{Pn2} yield
\bea
(1-v)\sum_{n=1}^\infty u_{n} P^{+-}_n v^{n-1} + \gamma ({\cal P}-{\cal G}) &=& 0 \label{calGg}\\
w (v-1) {\cal P} + (1-v) \sum_{n=1}^\infty u_{n} P^{++}_n v^{n-1} + \gamma ({\cal G}+{\cal H}-2 {\cal P}) &=& 0 \label{calPg}
\eea
As the equations \eq{Gn2}-\eq{Pn2} are defined in grand canonical ensemble, the constraint \eq{norm1} for $N \to \infty$ gives ${\cal G}(1)+{\cal H}(1)+2 {\cal P}(1)= \rho$, and from \eq{calHg} and \eq{calGg}, we find that in the stationary state, ${\cal G}(1)={\cal P}(1)={\cal H}(1)$. This result along with \eq{cons1} for $N \to \infty$ shows that for arbitrary $u_n$, the above equations for the generating functions need to be solved subject to the following constraints, 
\bea
{\cal G}(1)={\cal H}(1)={\cal P}(1)&=& \frac{\rho}{4} \label{cnst1} \\
{\cal G}'(1)+{\cal H}'(1)+2 {\cal P}'(1)&=& 1-\rho \label{cnst2}
\eea
where prime denotes the derivative w.r.t. $v$.

For the hop rate \eq{rate}, the equations \eq{calGg} and \eq{calPg} do not close, but we can write
\bea
(1-v) ({\cal G}-P^{+-}_0 +b S_G) +  \gamma v ({\cal P}-{\cal G}) &=& 0 \label{calGg2} \\
w v (v-1) {\cal P} + (1-v) ({\cal P}-P^{++}_0 +b S_P) + \gamma v ({\cal G}+{\cal H}-2 {\cal P}) &=& 0
\label{calPg2}
\eea
where $S_G(v) = \sum_{n=1}^\infty n^{-1} P^{+-}_n v^n$ and $S_P(v) = \sum_{n=1}^\infty n^{-1} P^{++}_n v^n$. On solving \eq{calGg2} for $S_G$ and differentiating both sides of the resulting equation, we obtain an expression for $S_G'$ which, by definition, obeys $v S'_G(v) = {\cal G}(v)-P^{+-}_0$. Thus, we arrive at a differential equation involving ${\cal G}', {\cal P}'$:
\bea
\frac{{\cal G}(v)-P^{+-}_0}{v} &=& \left(\frac{\gamma v ({\cal P}(v)-{\cal G}(v)) -(1-v) P_0^{+-}+(1-v)
  {\cal G}(v)}{b (v-1)} \right)'
\eea
 A similar  differential equation is obtained from \eq{calPg2} following these steps. 
 
 For convenience, using \eq{calHg}, we work with  ${\cal G}$ and ${\cal H}$, and obtain simultaneous, first order, inhomogeneous differential equations of the following form, 
\bea
{\cal G}' &=&c_1(v) {\cal G}+a_1(v) {\cal H}' +b_1(v) {\cal H}+i_1(v) \label{Gpcc}\\
{\cal H}' &=& c_2(v) {\cal H}+a_2(v) {\cal G}' +b_2(v) {\cal G}+i_2(v)  \label{Hpcc}
\eea
For $a_1, a_2 \neq 1$, the above equations can be reduced to the following form, 
\bea
{\cal G}' &=&\alpha_1(v) {\cal G}+\beta_1(v) {\cal H}+I_1(v) \label{Gpc} \\
{\cal H}' &=&\alpha_2(v) {\cal H}+\beta_2(v) {\cal G}+I_2(v) \label{Hpc}
\eea
On differentiating \eq{Gpc} w.r.t. $v$, and using \eq{Hpc} to eliminate ${\cal H}'$ in the resulting equation leaves an RHS involving ${\cal G}', {\cal G}, {\cal H}$. Then the generating function, ${\cal H}$ can be eliminated using \eq{Gpc}. 
We thus find that ${\cal G}$ obeys a linear, inhomogeneous, second order differential equation with variable coefficients,
\bea
&&{\cal G}''(v) -\frac{ \left(\alpha _1(v)+\alpha _2(v)\right) \beta _1(v)+\beta _1'(v)}{\beta _1(v)} {\cal G}'(v)\nn \\
&+&  \left[\alpha _1(v) \left(\alpha _2(v)+\frac{\beta _1'(v)}{\beta _1(v)}\right)-\alpha _1'(v)-\beta _1(v) \beta _2(v)\right] {\cal G}(v) \nn \\
&=& I_2(v) \beta _1(v)+I_1'(v)-I_1(v) \left(\alpha _2(v)+\frac{\beta _1'(v)}{\beta _1(v)}\right) \label{GO2}
  \eea 
Unfortunately, for $b > 0$, the coefficients of the above differential equation do not simplify and the model does not seem to be exactly solvable. 

For completeness, the coefficients in \eq{Gpcc} and \eq{Hpcc} are given below: 
\bea
a_1 &=& \frac{v (\gamma -v w+w)}{\gamma  v+v-1} \\
b_1&=& \frac{\gamma +(v-1)^2 w}{(1-v) (\gamma  v+v-1)} \\
c_1&=& \frac{b (v-1)^2-\gamma  v}{(1-v) v (\gamma  v+v-1)} \\
i_1 &=& \frac{b (v-1) P_0^{+-}}{v (\gamma  v+v-1)} \\
a_2 &=& \frac{\gamma ^2 v}{\gamma ^2 v-\gamma  (v-1) (3 v w-1)+(v-1)^2 w (v w-1)}\\
b_2 &=& -\frac{\gamma ^2}{(v-1) \left(\gamma ^2 v-\gamma  (v-1) (3 v w-1)+(v-1)^2 w (v w-1)\right)}\\
c_2 &=& \frac{v \left(\gamma ^2+3 \gamma  (v-1)^2 w+(v-1)^2 w (-2 v w+w+1)\right)-b (v-1)^2
   (\gamma -v w+w)}{(v-1) v \left(\gamma ^2 v-\gamma  (v-1) (3 v w-1)+(v-1)^2 w (v
   w-1)\right)}\\
i_2 &=& \frac{b \gamma  (v-1) P_0^{++}}{v \left(\gamma ^2 v-\gamma  (v-1) (3 v w-1)+(v-1)^2 w (v
   w-1)\right)}
\eea

\section{Persistent exclusion process}
\label{app_pep}

To understand the behavior of mean mobility in the fluid phase, here we consider the model when the hop rate is constant in $n$.  On setting $b=0$ in \eq{calGg2} and \eq{calPg2}, we obtain a closed set of equations for the generating functions that can be easily solved. We find that  
\bea
{\cal H}(v) &=& \frac{\gamma P^{++}_0}{(1+\gamma) v_r w^2}~\frac{v-v_r}{(v-v_0) (v-v_+) (v-v_-)}
\eea  
where,
\bea
v_r &=& \frac{P^{++}_0}{\gamma (P^{++}_0+P^{+-}_0)+P^{++}_0}\\
v_0 &=& \frac{\gamma+w}{(1+\gamma) w} \\
v_\pm &=& \frac{2 \gamma +w +1 \pm \sqrt{\left(2 \gamma +w +1\right)^2-4 w }}{2 w }
\eea
Similarly, explicit expressions for ${\cal G}$ and ${\cal P}$ can be obtained using 
\bea
{\cal P}(v) &=& \frac{\gamma+(1-v) w}{\gamma} {\cal H}(v) \\
{\cal G}(v) &=& \frac{ v (\gamma -v w+w) {\cal H}(v)+P_0^{+-} (v-1)}{\gamma  v+v-1}
\eea

The three unknowns, namely, $P^{++}_0, P^{+-}_0, w$ in the above equations can be obtained as follows. From the normalization condition \eq{cnst1}, we find that 
\bea
\frac{P^{+-}_0+P^{++}_0}{1-w } &=& \frac{\rho}{2} 
\eea
which also shows that $w  < 1$. Since $v_- < 1$ while $v_0, v_+ > 1$, the pole at $v=v_-$ in the above generating functions implies that the corresponding distributions increase exponentially for large $n$. 
Then using the conservation condition \eq{cnst2} and demanding that  $v=v_- < 1$ be a zero of the generating functions, we find that the mean mobility is determined by the following equation, 
\bea
\frac{\rho-4 \gamma   \left(\rho +w -1\right)-\rho  w  \left(3 w +2\right)}{4 \gamma   \left(\rho +w -1\right)+\rho  \left(w +1\right)^2}=\frac{1-w +\sqrt{\left(2 \gamma  +w +1\right)^2-4 w }}{2 \left(\gamma  +w +1\right)} \label{exactw}
 \eea 
On inverting the generating functions, following exact expressions for the distributions are also obtained:
\bea
P^{-+}_n &=& \frac{C \left(v_+^{-n-1} - v_0^{-n-1}  \right)}{v_0-v_+}  \label{appPmpb0}\\
P^{++}_n &=& \frac{C [(\gamma +w) \left(v_+^{-n-1}-v_0^{-n-1}\right)+w \left(v_0^{-n}-v_+^{-n}\right)]}{\gamma  \left(v_0-v_+\right)}
\eea
where,
\be
C=\frac{\gamma  \left(\rho-2 \gamma  \left(\rho  w^2+\rho +2 w-2\right)-\rho  w (3 w+2)
   \right)}{4 (\gamma +1) w^2 (w+1)}
\ee
Using these in \eq{Gn2}, $P^{+-}_n$ can also be found. As expected, all the distributions decay exponentially with the size of the hole cluster.

From the exact expression \eq{exactw} for the mean mobility, we find that the qualitative behavior of the mobility with switching rate and density is similar to that in the fluid phase for $b=2.5$ which is shown in Fig.~\ref{mobi}. For $\gamma  \gg 1$, \eq{exactw} shows that 
\be
 w =1-\rho-\frac{\rho (1-\rho) (2-\rho)}{2 \gamma }+... \label{wb0larg}
 \ee
 which matches \eq{p0fug} for $\gamma \to \infty$ and the correction term is in agreement with that obtained in Sec.~\ref{pertL}. Equation \eq{exactw} also shows that 
 \bea
 w  &\stackrel{\gamma  \to 0, \rho~arbit}{\approx}& \frac{4 \gamma  (1-\rho)}{\rho} , \gamma  < \frac{\rho}{4 (1-\rho)} \label{wapp1}\\
w  &\stackrel{\rho \to 0, \gamma  ~arbit}{\approx}& 1-\frac{\rho(1+\gamma)}{\gamma},  \rho < \frac{\gamma}{1+\gamma}  \label{wapp2}
 \eea
 so that the two limits do not commute, as discussed in the main text. But in the scaling limit, $\rho, \gamma  \to 0, w \stackrel{s=\frac{\rho}{\gamma }}{\to} w_s$, from \eq{exactw}, we find that
 \bea
 s w_s^3+2 s w_s^2+(s+4) w_s-4=0  \label{wscube}
 \eea
 This cubic equation has two complex roots and one real root, but it is easy to see that 
 \bsn
 {w_s \approx  \label{wb0sl}}
 1-s~,~s \ll 1 \label{wb0s}\\
  \frac{4}{s}~,~s \gg 4 \label{wb0l}
 \esn
 As $w_s \stackrel{s \to \infty}{\longrightarrow} 0$, keeping $s w_s$ finite in \eq{wscube}, we immediately obtain  \eq{wb0l};  similarly, since $w_s \stackrel{s \to 0}{\longrightarrow} 1$, on writing the equation for $1-w_s$, we are led to \eq{wb0s}. 
{Note that \eq{wapp1} and \eq{wapp2} reduce to \eq{wb0sl}  on taking the appropriate limits.}

We also mention that on expanding $C, v_0, v_\pm$ for $\gamma \to 0$, from \eq{appPmpb0}, we obtain
\bea
P_n^{-+} \approx \frac{\gamma ^2 \left(4+s-3 s w_s^2-2 (s+2) w_s\right) \left(1-w_s^{n+1}\right)}{4
   \left(1-w_s\right) w_s \left(w_s+1\right)}
\eea   
which is of order $\gamma^2$, as argued in Sec.~\ref{pertS}. Furthermore, on using \eq{wb0sl}, we find that $P_n^{-+} \propto \gamma^2 s^2 \propto \rho^2$ for both small and large $s$.

\section{Mobility for large switching rates}
\label{app_pertL}

From the discussion in Sec.~\ref{pertL} for large $\gamma$, we find that the first order correction $p_n^{++}$ obeys the following equation,
\bea
u_{n+1} p^{++}_{n+1}-  \left( u_n+{\cal w} \right) p^{++}_n + {\cal w} p^{++}_{n-1}=q_n, n \geq 0 \label{app_Pinhomo}
\eea
where, $u_0=p^{++}_{-1}=0$ and
\bea
q_n &=& - \frac{w_1}{4} ({\cal K}_{n-1}-{\cal K}_n) -\frac{{\cal w}^2}{8} ({\cal K}_{n-2} -2 {\cal K}_{n-1} +{\cal K}_n) \nn \\
&-& \frac{{\cal w}}{8} [u_{n+1} ( {\cal K}_{n+1}- {\cal K}_{n}) - u_n   ({\cal K}_{n}-{\cal K}_{n-1} )] \label{app_qn}
\eea
with ${\cal K}_{-2}={\cal K}_{-1}=0$.
It can be verified that for $n \geq 2$, the  general solution of the inhomogeneous equation \eq{app_Pinhomo} is given by
\bea
p^{++}_n &=& c_1 {\cal K}_n+c_2 {\cal K}_n \sum_{k=2}^n \frac{{\cal K}_0}{{\cal K}_{k-1}}+\frac{{\cal K}_n}{{\cal w}} \sum_{k=2}^n \frac{1}{ {\cal K}_{k-1}} \sum_{m=1}^{k-1}  q_m \\
&=& c_1 {\cal K}_n+  {\cal K}_n \sum _{k=2}^n \left(\frac{ u_{k-1}}{8 
   }+\frac{ u_k}{8  }-\frac{{\cal w}}{4}+\frac{w_1}{4 {\cal w}}\right) \nn \\
&+&  \left(c_2+ \frac{{\cal w}}{4}-\frac{u_1}{8}-\frac{w_1}{4 {\cal w}}\right) {\cal K}_n \sum _{k=2}^n \frac{{\cal K}_0}{{\cal K}_{k-1}} 
~,~n \geq 2 \label{Pn12}
\eea
where $c_1$ and $c_2$ are constants. Using (\ref{app_Pinhomo}) for $n=1, 2$ and the above result, the probabilities $p^{++}_0$ and $p^{++}_1$ can also be found. 

To find the three unknowns, {\it viz.}, $c_1, c_2, w_1$ in the above equation, we use  the boundary condition ($p^{++}_{-1}=0$) in \eq{app_Pinhomo} for $n=0$, the normalization condition \eq{norm1} and the conservation equation \eq{cons1} for infinitely large system that, respectively, yield
\bea
u_{1} p^{++}_{1}- w_0  p^{++}_0 &=& q_0 \label{bcg} \\
\sum_{n=0}^\infty (2 p_n^{++}+p_n^{+-}+p_n^{-+})=4 \sum_{n=0}^\infty p_n^{++} &=& 0 \label{constr1}\\
\sum_{n=0}^\infty n (2 p_n^{++}+p_n^{+-}+p_n^{-+}) =4 \sum_{n=1}^\infty n p_n^{++} &=& 0 \label{constr2}
\eea
on using $p_n^{+-}+p_n^{-+}=2 p_n^{++}$ due to (\ref{Gfeqn}). After some simplifications, from (\ref{bcg}), we obtain
\bea
w_1 &=& -\frac{{\cal w}}{2} \left(1+b-8 c_2 -2 {\cal w}\right) \label{z1final}
\eea
which renders the last term on the RHS of  (\ref{Pn12}) to zero. For $n \geq 2$, we thus have 
\bea
p_n^{++} &=& \left[c_1+\frac{n}{8} \left(1-b+8 c_2\right)+\frac{b}{8 n}+\frac{b}{4} (\psi(n)+\gamma_{e}
   )-\frac{8 c_2+1}{8}  \right] {\cal K}_n \label{Pn13}
\eea   
where $\gamma_{e}$ is the Euler's constant and $\psi(n)=\frac{\Gamma'(n)}{\Gamma(n)} \stackrel{n \gg 1}{\sim} \ln n$ is the polygamma function. 
Using (\ref{Pn13}) in the constraint equations (\ref{constr1}) and (\ref{constr2}), the coefficients $c_1$ and $c_2$ can be found, which finally yields the correction to mean mobility through \eq{z1final}. 

The above analyses is valid only when the density is in the fluid phase of the passive model; 
this is because for $\rho \leq {\cal r}_c$ where the passive mobility ${\cal w}=1$, equation \eq{Pn12} and hence the simultaneous equations (\ref{constr1}) and (\ref{constr2}) for $c_1, c_2$ are independent of density, resulting in constant $w_1$. However, as Fig.~\ref{mobi} shows, the active critical density $\rho_c < {\cal r}_c$, and therefore, the mobility varies  with density for $\rho_c < \rho < {\cal r}_c$. 
Thus the perturbation theory to linear order in ${\gamma}^{-1}$ shows that the maximum density where the mobility  is independent of $\rho$ is ${\cal r}_c$, and we expect the true critical density to be of quadratic or higher order in ${\gamma}^{-1}$.

\section{Mobility for small switching rates}
\label{app_pertS}

For $\rho, \gamma \to 0$ with $s=\frac{\rho}{\gamma} \gg 1$, the hole cluster distributions in the stationary state are governed by \eq{G0sca}-\eq{Pnsca}. Using these equations, we find that the corresponding generating functions defined in \eq{genfns} obey the following approximate equations, 
\bea
{\cal G}(v) &\approx& \frac{v^b I(v)-b \int_0^v z^{b-1} I(z) dz+ v^b P^{+-}_0}{v^b} \label{calGsm} \\
{\cal H}(v) &\approx& \frac{\gamma {\cal P}(v)}{w (1-v)} \label{calHsm} \\
{\cal P}(v) &\approx& g(v w) P^{++}_0 
\eea
where, $g(x)$ is given by \eq{gdefn}, and 
\bea
I(v)=\frac{(1-v+\gamma v) P^{+-}_0}{1-v} -\frac{\gamma v {\cal P}(v)}{1-v} 
\eea
The generating function ${\cal G}$ obeys $(1-v) ({\cal G}-P^{+-}_0 +b S_G) +  \gamma v ({\cal P}-P^{+-}_0) \approx 0$ where, $S_G$ is defined in Appendix~\ref{app_arbb}. For $b > 0$, taking a derivative of this equation w.r.t. $v$, carrying out the integration, and noting that the constant of integration must vanish to prevent exponentially diverging distribution, we finally obtain \eq{calGsm}. 

To determine $w, P_0^{++}, P_0^{+-}$ in the above expressions, we invoke the normalization and conservation constraints given by \eq{cnst1} and \eq{cnst2}. We first note that \eq{calHsm} and ${\cal H}(1)={\cal P}(1)$ give $w=\gamma/\epsilon$ where $\epsilon=1-v$. But since the mobility is finite, $\gamma \to 0, v \to 1$ such that $\gamma/\epsilon$ remains finite. Then, on using \eq{cnst1} for ${\cal G}$ and ${\cal P}$,  
the probabilities $P_0^{++}, P_0^{+-}$ can be written in terms of $\epsilon$. Finally, using the above results in \eq{cnst2}, we find that $\epsilon$ is a solution of the following equation:
\bea
&& \frac{b \left(b \epsilon ^2-\epsilon  \gamma +\gamma \right) (\gamma  \rho  B_{1-\epsilon }(b+1,0)-4
  I_G(1-\epsilon ))}{(\epsilon -1) \left((1-\epsilon )^b (\epsilon  (\gamma -1)-\gamma )+b \epsilon 
   \gamma  B_{1-\epsilon }(b+1,0)\right)} 
+\frac{\rho  (\gamma -\epsilon  (\gamma +3)) \, _2{F}_1\left(1,2;b+1;\left(\frac{1}{\epsilon }-1\right)
   \gamma \right)}{(\epsilon -1) \, _2{F}_1\left(1,1;b+1;\left(\frac{1}{\epsilon }-1\right) \gamma \right)} \nn \\
&=& \frac{\rho  \left(4 \epsilon ^2+(\epsilon -1) \gamma -1\right)}{1-\epsilon}+4 \epsilon \label{eptrns}
\eea
where, $B_z(x,y)=\int_0^z dt \; t^{x-1} (1-t)^{y-1}$ is the incomplete beta function and  $I_G(v)=\int_0^v dz\; \frac{\gamma z^b}{1-z} {\cal P}(\frac{z \gamma}{\epsilon})$. 

For $b=0$, \eq{eptrns} simplifies to yield the following quadratic equation for $\epsilon$, 
\bea
4  (\gamma +1) (\rho -1) \epsilon ^2+  (\gamma  (\gamma  \rho +4)+\rho ) \epsilon-\gamma  (\gamma +1) \rho = 0 \label{scab0e}
\eea
This is unlike in Appendix~\ref{app_pep}, where the mobility obeys the cubic equation \eq{wscube}. However, for $\rho, \gamma, \epsilon \to 0$ with $\gamma/\rho$ and $\gamma/\epsilon$ finite, from \eq{scab0e}, we obtain $4 \epsilon ^2-\epsilon  (4 \gamma +\rho )+\gamma  \rho \approx 0$ that yields $w=1$ and $4/s$. The latter solution is acceptable as  
the theory in Sec.~\ref{pertS} is valid for perturbation about the zero mobility state, and is also in agreement with \eq{wb0l}. 

For $b > 0$, using $B_{1-\epsilon }(b+1,0) \stackrel{\epsilon \to 0}{\approx} -\ln \epsilon$ and $\, _2{F}_1\left(a_1,a_2;a_3; z\right) \stackrel{z \to 0}{\approx} 1+\frac{a_1 a_2}{a_3} z$, we find that $I_G(v) \stackrel{\epsilon \to 0}{\approx} -\frac{\gamma \rho}{4} \ln\epsilon$ and \eq{eptrns} reduces to the following equation:
\bea
4 (b+1)  (\rho -1) \epsilon ^2+  \rho  \left(b+\gamma ^2+3 \gamma +1\right) \epsilon-\gamma ^2 \rho \approx 0
\eea
For small $\rho, \epsilon, \gamma$, the above equation simplifies to yield $-4 (b+1) \epsilon ^2+\rho (b+1) \epsilon \approx 0$ whose roots are given by $0$ and $\rho/4$ so that the latter root yields $w \approx 4/s$, which is independent of $b$ and is the same as for $b=0$.

\clearpage

\end{document}